\documentclass[a4paper,11pt]{article}
\usepackage[utf8]{inputenc}
\usepackage[italian,english]{babel}
\usepackage{braket}
\usepackage{amssymb}
\usepackage{amsfonts}
\usepackage{amsmath}
\usepackage{fancyhdr}
\usepackage{lineno}
\usepackage{fullpage}

\newcommand{\bnabla}{\bar\nabla}
\def\rgb{\sqrt{ - \bar g }}

\usepackage{subfigure}

\def\lt{\left(}
\def\rt{\right)}

\newcommand{\rg}{\sqrt{g}}

\newcommand{\de}{\delta}
\newcommand{\pa}{\partial}
\newcommand{\nn}{\nonumber\\}
\newcommand{\lsim}{\mathrel{\mathop{\kern 0pt \rlap
  {\raise.2ex\hbox{$<$}}}
  \lower.9ex\hbox{\kern-.190em $\sim$}}}
\newcommand{\gsim}{\mathrel{\mathop{\kern 0pt \rlap
  {\raise.2ex\hbox{$>$}}}
  \lower.9ex\hbox{\kern-.190em $\sim$}}}
\newcommand{\dfun}[2]{ \frac{\delta #1}{\delta #2}}

\newcommand{\be}{\begin{equation}}
\newcommand{\ee}{\end{equation}}
\newcommand{\beqa}{\begin{eqnarray}}
\newcommand{\eeqa}{\end{eqnarray}}
\newcommand{\bea}{\begin{eqnarray}}
\newcommand{\eea}{\end{eqnarray}}

\newcommand{\beq}{\begin{equation}}
\newcommand{\eeq}{\end{equation}}

\newcommand{\sm}{\mathcal{S}}

\usepackage{pstricks}
\usepackage{color}
\usepackage{multirow}

         \let\d=\delta
         
        \let\m=\mu
\let\n=\nu      \let\x=\xi           
        
     \let\y=\psi    
\let\D=\Delta
\let\d=\delta

\newcommand{\sdfrac}[2]{\mbox{\small$\displaystyle\frac{#1}{#2}$}}

\newcommand{\vf}{\phi}

\newcommand{\sq}{\Box}

\newcommand{\id}{\mbox{1 \kern-.59em {\rm l}}}

\begin{document} 

\begin{center}
	\vspace{1.5cm}
	{\Large \bf $4d$ Einstein Gauss-Bonnet Gravity without a Dilaton \footnote{Proceedings of the Corfu Summer Institute 2022 "Workshops on the Standard Model and beyond" }\\}
\vspace{0.3cm}
{\bf $^{(a)}$Claudio Corian\`o, $^{(a)(b)}$Mario Creti, ${}^{(a)}$ Stefano Lionetti, $^{(c)}$Matteo Maria Maglio, $^{(a)}$Riccardo Tommasi\\}
		\vspace{1cm}
		{\it  $^{(a)}$Dipartimento di Matematica e Fisica, Universit\`{a} del Salento \\
		and INFN Sezione di Lecce, Via Arnesano 73100 Lecce, Italy\\
		National Center for HPC, Big Data and Quantum Computing\\}
	\vspace{.5cm}
	{\it  $^{(b)}$Center for Biomolecular Nanotechnologies, Istituto Italiano di Tecnologia, Via Barsanti 14, 73010 Arnesano, Lecce, Italy.
		\\}
\vspace{.5cm}
{\it  $^{(c)}$ Institute for Theoretical Physics (ITP), University of Heidelberg\\
	Philosophenweg 16, 69120 Heidelberg, Germany} 

\end{center}

\begin{abstract}
A nonlocal version of  Einstein-Gauss Bonnet (EGB) gravity can be generated by a procedure that follows closely the steps of the derivation of the conformal anomaly effective action. The action is obtained using a Weyl decomposition of the metric, with the conformal factor removed by a finite renormalization of the topological term. We outline the mains steps in the derivation of this action, stressing on the analogies and differences respect  to the anomaly action and to the ordinary $4d$ EGB theory formulated as a special version of dilaton gravity. These formulations allow a systematic investigation of the nonlocal $R\Box^{-1}$ corrections to General Relativity. In the case of conformal anomaly action they are motivated by the generation of a conformal backreaction, due to the breaking of conformal symmetry in the early universe. 
\end{abstract}

\section{Introduction}
Modifications of General Relativity (GR) may include terms of higher orders in the Riemann tensor and its contractions, if confined just to the metric tensor, or may combine additional fields beside the metric, such as extra scalars. These corrections are typically introduced to account for phenomena that are not adequately described by classical GR alone, such as the behavior of spacetime at extremely high energies or in the presence of strong gravitational fields. \\
There have been various proposals for quadratic corrections to GR, each with its own motivations and limitations, most of them formulated as effective actions generated by the integration of a quantum matter sector in the functional integral. 
Other modifications may include higher-order derivative terms in the gravitational action. 
The inclusion of these terms is motivated by attempts to address the renormalizability of gravity \cite{Stelle:1977ry}. \\
Among the quadratic corrections, of particular interest are those incorporating the Gauss-Bonnet term, that arises in certain higher-dimensional theories of gravity. It involves the contraction of the Riemann curvature tensor with itself and is a topological invariant in four dimensions. Gauss-Bonnet gravity has been studied extensively in the context of braneworld scenarios and string theory \cite{Zwiebach:1985uq}.\\
Another interesting class of corrections are those that modify the gravitational field equations with the inclusion of $f(R)$ terms in the action, where the Einstein-Hilbert action is replaced by a general function of the Ricci scalar curvature. Quadratic corrections can arise naturally in these modified gravity frameworks, for instance starting from the simple $R + R^2$ theory \cite{Starobinsky:1980te} 
\begin{equation}
\label{ST}
\mathcal{S}_{Starobinsky}=\frac{1}{2 \kappa}\int \!d^4x \sqrt{-g}\,\bigl[R+\alpha R^2\bigr]+\mathcal{S}^{(m)} ,
\end{equation}
(see for instance the discussions in \cite{Antoniadis:2020dfq,Capozziello:2021krv,Kanti:2015pda}),
where $\mathcal{S}^{(m)}$ is the ordinary matter action. The quadratic term $R^2$ can give rise to the  acceleration of the early universe with an inflationary behavior. \\
It's worth noting that while quadratic corrections to GR offer interesting avenues for exploration and can address certain shortcomings of the theory, they also come with their own challenges. These include the need to ensure consistency with experimental tests, such as solar system observations and gravitational wave detections, as well as theoretical constraints from principles like causality and energy conditions. \\
In the context of higher derivative gravity theories,  the issue of unitarity and the potential presence of ghosts becomes a significant concern.
For example, in Lovelock theory \cite{Lovelock:1971yv} \cite{Charmousis:2014mia}, the specific combination of higher curvature terms leads to a consistent theory without ghosts.
However, in general, the introduction of higher derivative terms increases the complexity of the theory and makes it more challenging to maintain unitarity. Ghosts can arise if the coefficients or combinations of these higher derivative terms are not properly chosen. \\
The Einstein-Gauss-Bonnet theory (EGB), proposed by Lanczos \cite{Lanczos:1938sf,Lanczos:1932zz} and generalized by Lovelock \cite{Lovelock:1971yv}, holds a special position among alternative theories to GR. What sets it apart is that it requires no additional fundamental fields beyond those already present in GR, while ensuring that its field equations can be expressed with no more than second derivatives of the metric. This is crucial because higher-derivative theories can lead to the Ostrogradsky instability 
(see \cite{Woodard:2015zca}), which can result in unphysical behavior. As a result, the theory is highly motivated and occupies an important position in the landscape of possible alternatives to GR.\\
In $d=4$ the GB term is topological, and this has been the main reason to discard such contribution in realistic cosmology.\\
 However, a recent study \cite{Glavan:2019inb} has suggested that a nontrivial contribution to the Gauss-Bonnet (GB) term in a 4-dimensional (4d) theory can be derived by taking the $d\to 4$ limit of the fields while simultaneously performing a singular limit on the coupling. However, this proposal poses several issues as it requires a specific compactification to be consistently formulated. In a general compactification, the extra degrees of freedom of the metric manifest as extra fields, and assumptions must be made regarding their dependence on the extra dimensions as we move to $d=4$ \cite{Lu:2020iav,Matsumoto:2022fln,Aoki:2020lig,Hennigar:2020lsl,Gurses:2020ofy}. For example, in compact extra dimensions, a harmonic expansion generates an infinite tower of Kaluza-Klein modes that must be taken into account. One possibility is to consider only the zero mode in the classical d=4 theory, which is known as dimensional reduction (DRED). This procedure neglects the dependence of the fields on the extra dimension and is part of the regularization process of the effective action at $d=4$, which includes extra scales in the resulting Lagrangian \cite{Coriano:2022ftl}. Despite these challenges, the interest in this type of $4d$ theory has been remarkable, as it allows evading the strict requirements of Lovelock's theory. As just mentioned, this requires a rescaling of the coupling constant 

\beq
\alpha \to \frac{\alpha}{(d-4)}.
\eeq
In \cite{Glavan:2019inb}, it was suggested that by incorporating a rescaling into the Lanczos tensor, the terms involving the renormalized coupling could remain finite and non-vanishing, resulting in a regular $0/0$ limit. If successful, this approach would enable the Gauss-Bonnet term to directly impact the $4d$ theory of gravity without requiring the inclusion of extra degrees of freedom. Although not particularly new from the perspective of regulating quantum corrections in the conformal anomaly effective action, the interest of the GR community in this extension has been remarkable. The regularization of these theories follows the ordinary Wess-Zumino method \cite{Mazur:2001aa,Coriano:2022ftl}, which involves the Weyl rescaling of the metric for the counterterms $V_E$ and $V_{C^2}$, which are defined below. The regularization of the resulting action requires a subtraction, which can be performed in several ways due to the arbitrariness of finite subtractions.
\subsection{Conformal backreaction}
Modifications of gravity by the inclusion of a quantum matter sector may lead to very interesting realizations if the matter sector satisfies specific spacetime symmetries. 
For example, there are some important implications for the structure of the effective action if the matter sector is conformal \cite{Codello:2012sn,Coriano:2013xua,Coriano:2013nja}. \\
The integration of a conformal sector allows to derive a form of gravity which is expressed uniquely in terms of corrections extracted from the two invariants $V_E$ and $V_{C^2}$, defined in terms of the Euler density and the square of the Weyl tensor

\begin{align}
\label{ffr}
V_{E}(g,d)\equiv &\mu^{\epsilon} \int\,d^dx\,\sqrt{-g}\,E \notag \\ 
V_{C^2}(g, d =4)\equiv & \mu^{\epsilon}\int\,d^4x\,\sqrt{-g}\, C^2, 
\end{align}
where

\begin{align}
C^{(d) \alpha\beta\gamma\delta}C^{(d)}_{\alpha\beta\gamma\delta}
&=
R^{\alpha\beta\gamma\delta}R_{\alpha\beta\gamma\delta} -\frac{4}{d-2}R^{\alpha\beta}R_{\alpha\beta}+\frac{2}{(d-2)(d-1)}R^2\label{Geometry1}
\end{align}
is the square of the Weyl tensor. We will neglect the dependence on the scale $\mu$ in our discussion. 
In general this dependence is accompanied also by the extra dimensional scale $L$ in the integration over the 
$d-4$ coordinates of the metric, which are neglected in DRED.
In four dimensions, $E$, appearing in  \eqref{ffr}, is proportional to the Euler characteristic of the spacetime, identifying its topological structure. In this second case the tensor indices in its definition run from 0 to 3 and its original expression reduces to 
\beq
E\equiv R_{\mu\nu\alpha\beta}R^{\mu\nu\alpha\beta}-4R_{\mu\nu}R^{\mu\nu}+R^2.  
\eeq
The general expression given in \eqref{ffr}, with indices varying from $0$ to $d-1$ define its extension to $d$ dimensions.\\
Of these two terms, only the first one is present in the $4d$ EGB 
theorie(s), which are purely classical variant of the renormalized anomaly action. They contain some tracts of the conformal anomaly action, as discussed in \cite{Coriano:2022ftl,Coriano:2023lmc}, but also differences, for being obtained from a singular classical limit on the coupling of the GB term, accompanied by the reduction of the metric to $d=4$. \\
As we are going to see, the limit is not uniquely identified. 
The approach borrows on the standard formalism of dimensional regularization (DR) of an effective action in field space, with finite subtractions which depend on the prescription. For example, in a conformal anomaly action, logarithmic terms are automatically generated in its expression if these subtractions are defined quite closely to what is usually done in flat space. We have identified two types of subtractions in\cite{Coriano:2022ftl}, which will be commented upon in the next sections.\\
 On the other hand, the elimination of the dilaton field in the nonlocal version of such actions is directly related to the redefinition of $V_E$. If we restrict ourselves to the case of a $4d$ GB theory, which is defined only by the inclusion of a GB term, without any quantum correction, the generation of a nonlocal GB theory, for example, is entirely associated with the addition of a  $\epsilon R^2/18$ term in $V_E$ in dimensional regularization (DR), with 
$\epsilon=d-4$. \\
These points have been discussed in the $4d$ EGB case in \cite{Coriano:2022ftl,Coriano:2023sab}. 
For instance, the elimination of the dilaton (conformalon) field in the generation of nonlocal versions of such theories, discussed in the case of anomaly induced actions, plays a crucial role in their derivation. In general, however, the procedure needs to be amended as we move towards the 
flat spacetime limit, since nonlocal actions of this type have been shown to be in disagreement with the perturbative effective action, if  this is computed in the standard diagrammatic approach with Feynman diagrams. The disagreement holds beyond 3-point functions, as shown in \cite{Coriano:2022jkn}. \\
If conformal symmetry is bound to play an important role in the dynamics of the early universe, then such actions and their nonlocalities need to be studied with care. Nonlocal modifications of GR  are at the center of several investigations, also in connection with the production of gravitational waves, as in 
\cite{Belgacem:2017cqo,Capozziello:2021bki}. \\
Such nonlocal actions, connected with the trace anomaly, are specific, and are characterised by the presence of multiple insertions of $1/\Box$ interactions, corresponding to anomaly poles.  They may provide a reference point for the interesting nonlocal extensions discussed in the recent literature \cite{Capozziello:2021krv}, for being associated with the quantum breaking of an important symmetry. \\ 
 Here, we are going to review these aspects in some detail, in order to clarify some of the technical issues discussed in our previous works. 
 
 \section{ DR with Dimensional Reduction (DRED)}
To reduce the effective action from $d$ dimensions to $d=4$, we need two scales: the renormalization scale $\mu$ to regulate the UV behavior of the theory, and the IR scale $L$ from the extra dimensions. This applies to both classical and quantum cases, and the dependence on these scales is logarithmic, affecting $V_{C^2}$. Neglecting the extra dimensional components of the metric leads to the appearance of the second scale after dimensional reduction. The resulting action in $d=4$ takes the form of dilaton gravity, which naturally exhibits scale breaking. When a conformal matter sector is coupled to a gravitational theory, it can backreact on the gravitational metric, and the reduction to $d=4$ naturally breaks conformal symmetry. However, this is not taken into account in the local $4d$ EGB presented in recent literature, as the $d\to 4$ limit is performed with respect to a fiducial metric in $d$ dimensions ($\bar g$), which misses finite terms containing a logarithmic scale. Further details and examples of how to handle these cutoffs are provided in the following sections and in the appendix.
Notice that of the two counterterms $V_E$ and $V_{C^2}$ introduced in the renormalization of a quantum conformal sector, only the second one is necessary for the cancelation of the singularities coming from the diagrammatic expansion, and it is nontopological. The first, instead, is introduced in order to satisfy the Wess-Zumino consistency condition in the effective action \cite{Coriano:2023sab}. This term is evanescent at $d=4$. The evanescence of 
vertices associated with $V_E$ has been discussed recently in \cite{Coriano:2022jkn}, where we have presented a method for the extraction of the minimal number of form factors of such effective actions, due to the presence of tensor degeneracies related to Lovelock identities \cite{Edgar:2001vv,lovelock_1970}. \\

\subsection{Special features of the GB term} 
Before delving into a comprehensive examination of the quantum effective action, it is important to discuss some notable characteristics of the Gauss-Bonnet term ($V_E$) that have garnered considerable attention in various contexts, including string theory. One particularly remarkable feature is its ability to eliminate double poles in the propagators of specific linear combinations of higher derivative invariants within the action.
 In general, expanding the Riemann tensor in fluctuations $h_{\mu\nu}$ around a flat Minkowski vacuum 

\beq
R_{\mu\nu\rho\sigma}=R^{(1)}_{\mu\nu\rho\sigma} + R^{(2)}_{\mu\nu\rho\sigma} +\ldots,
\eeq
shows quite clearly that at quadratic level the theory, arrested at the $"R^2"$ term, is plagued by propagating double poles, corresponding to the kinetic operator $\Box^2$, as one can easily derive from the action
\beq
\sm_2=\int d^d x \left( (R_{\mu\nu\rho\sigma})^2 + a (R_{\mu\nu})^2 + b R^2 \right),
\eeq
expanded to quadratic order in the fluctuations $h$. In the harmonic gauge ($\partial_\mu h^{\mu\nu}
=\frac{1}{2}\partial^{\nu}h$), with $h=h^{\mu}_\mu$, such double poles are associated with the action
\beq
\label{es1}
\sm_2^{(2)}=\frac{1}{4}\int d^d x \sqrt{g}\left(  (a+4) h^{\mu\nu}\Box^2 h_{\mu\nu} +(b-1) h\Box^2 h   \right), 
\eeq
 and vanish if we choose for $\sm$ the Gauss-Bonnet (GB) combination. 
Its contribution to the equation of motion appears through its first derivative with respect to the metric 

\begin{equation}
{V}_E ^{\mu\nu}=\frac{\delta}{\delta g_{\mu\nu}} V_E,
\end{equation} 

explicitly given by the relation 
\begin{equation}
V_E^{\mu\nu}= 4R_{\mu\alpha\beta\sigma}R^{\;\,\alpha\beta\sigma}_\nu-8R_{\mu\alpha\nu\beta}R^{\alpha\beta}-8R_{\mu\alpha}R^{\;\,\alpha}_{\nu}+4RR_{\mu\nu}-g_{\mu\nu}{E},
\end{equation}
vanishing at $d=4$ in any metric, as one can check. Notice that the variation of density $\sqrt{g} E$ with respect to the metric is a boundary distribution and not identically zero \cite{Coriano:2023sab}.\\
 The vanishing of $V_E^{\mu\nu}$, as shown by \eqref{es1}, is not expected to hold for a generic $d$. It is, instead, an obvious result at $d=4$, given the topological nature of the integral, for being proportional to the Euler number of the underlying spacetime manifold ($\mathcal{M}$)  
\beq
V_E=4 \pi \chi(\mathcal{M}). 
\label{top}
\eeq
$V_E$ shares properties similar to those of the EH action at $d=2$, 

\beq
\sm_{EH}(d)\equiv \int d^d x \sqrt{g} R,
\eeq
due to the topological nature of both functionals. \\
Clearly, all the functional derivatives of $V_E(4)$ or $\sm_{EH}(2)$ vanish for any 
metric $g_{\mu\nu}$ at $d=4$ and $d=2$ respectively,
\beq
\left(V_E(4)\right)^{\mu_1\nu_1\ldots \mu_n\nu_n}=0, \qquad \left(\sm_{EH}(2)\right)^{\mu_1\nu_1\ldots \mu_n\nu_n}=0, 
\eeq
while 
\beq
\label{ppf}
\left(V_E(d)\right)^{\mu_1\nu_1\ldots \mu_n\nu_n}\vert_{flat}=0, \qquad n=1,2.
\eeq
Eq. \eqref{ppf} shows that around flat space this term is responsible only for the generation of interacting classical vertices, from the trilinear level up. Therefore, the analysis of $V_E$ as a possible correction to the EH action, 

\beqa
\sm_{EGB}&=&\sm_{EH}(d) +\alpha V_E(d)\nn
&=& \frac{M_P^2}{2}\int d^d x \sqrt{g}\left(R + \frac{2}{M_P^2}\alpha E\right), 
\eeqa
giving equations of motion of the form 
\beq
G^{\mu\nu}  +\frac{2}{M_P^2}\alpha V_E^{\mu\nu}=0,\qquad G_{\mu\nu}=R_{\mu\nu} - \frac{1}{2}g_{\mu\nu} R, 
\eeq
is rather involved. As previously noted, it is essential to examine the term $\alpha V_E^{\mu\nu}$ thoroughly in the $d\rightarrow$ 4 limit. The final outcome of this analysis is contingent on the method employed to perform the limit, which leads to the emergence of extra degrees of freedom at $d=4$, as is typical of any compactification. Consequently, actions that depend on how the extra degrees of freedom of the metric are parameterized in the extra dimensional space arise as the limit is taken. \\
Purely gravitatonal actions generating second order equations of motion in any dimensions have been classified by Lovelock. We recall that Lovelock's theorem identifies the EH action - with the inclusion of a cosmological constant $\Lambda$ - as the unique, purely gravitational action, yielding second order equations of motion at $d=4$.  Obviously, the theorem does not contemplate in its assumptions the possibility of considering singular actions, regulated according to specific schemes, which is the case of 
\cite{Glavan:2019inb}. \\
\subsection{Lovelock}
We pause for a moment and briefly comment on the class of actions that generalize the GB term in  higher dimensions.\\
As just mentioned, in 1971 Lovelock demonstrated that in 3+1 dimensions, the only lagrangian theory with diffeomorphism invariance, second order equations of motion linear in the second derivatives of the metric, is the standard Einstein gravity with a cosmological constant. \\
Topological terms, in the form of the Euler densities, share this property in general dimensions 
\beq E_n = {1 \over 2^{d/2}} \ \de^{\nu_1 \nu_2 ... \nu_d}_{\mu_1 \mu_2 ... \mu_d} \
R^{\mu_1 \mu_2}_{\qquad \nu_1 \nu_2} R^{\mu_3 \mu_4}_{\qquad \nu_3 \nu_4} \ ... \ R^{\mu_{d-1} \mu_d}_{ \qquad \nu_{d-1} \nu_d},\eeq
with an action of the form
\beq S = \int d^dx \rg \sum_{n=0}^{[d/2]} \alpha_{2n} E_{2n} . \eeq 
Explicitly
\beqa
E_0 &=& 1 ,\nn
E_2 &=& R ,\nn
E_4 &=& R^{\mu \nu \rho \sigma} R_{\mu \nu \rho \sigma} - 4 R^{\mu \nu}R_{\mu \nu} + R^2,
\eea
and so on for cubic and for terms of higher orders.\\
In the variant proposed in \cite{Hennigar:2020lsl,Fernandes:2020nbq,Lu:2020iav}, the regulated action extends to the GB term an approach that has been introduced in the past in $d=2$ for the EH action. 
A careful analysis shows that the results of this procedure is to generate actions of Horndeski type, which are characterised by equations of motion of second order, but with the inclusion of a dilaton field. As such they correspond to a special form of dilaton gravity rather than to pure gravitational theories. 
Since the  steps introduced derivation of the $4d$ EGB action are contained in the analysis of the with the  anomaly actions, we turn to a discussion of the latter.
\section{From the conformal anomaly actions to the $4d$ EGB theory}
As discussed above, Einstein-Gauss-Bonnet (EGB) theories in four dimensions (4d) are obtained by taking a singular limit on the coupling constant of the topological Gauss-Bonnet term. Although this term has no dynamical content in $4d$, its coupling can be infinitely renormalized to obtain a finite action. By exploiting its evanescence in the equations of motion of the metric, one can derive a $(0/0)$ contribution to the classical action, that includes both gravity and a dilaton field. This allows for the definition of a quartic dilaton gravity theory, extensively studied in the previous literature and reviewed in \cite{Fernandes:2022zrq}.
However, there are other possibilities.\\
Drawing from previous work on anomaly actions and the inclusion of a finite renormalization of the topological Gauss-Bonnet term, one can show that a $4d$ EGB action  quadratic in the dilaton field can be expressed in a nonlocal form by solving for the same dilaton field in terms of the metric. This action has the same form as the one originally introduced by Riegert \cite{Riegert:1984kt} in the search for a functional solution of the anomaly constraint at $d=4$. \\
Riegert's analysis was not performed in the context of dimensional regularization, but an extension of this analysis was presented in \cite{Mazur:2001aa}, where the Wess Zumino (WZ) form of the action was discussed.\\
 Recently, we have shown that the WZ form is not the only possible one, since the regularization of $V_E$ and $V_{C^2}$ can be affected by finite subtractions associated with additional scales. In other words one can generate logarithms in the effective action if the subtraction terms are investigated quite closely to DR in flat space \cite{Coriano:2022ftl,Coriano:2023sab}. As we reduce the theory from generic dimensions to $d=4$ extra scales appear due to $V_{C^2}$, which is a true counterterm. \\
The action in \cite{Riegert:1984kt} was originally proposed as a way to find functional solutions to the anomaly constraint without considering Weyl invariant terms arising from the virtual corrections. As a result, the additional logarithmic terms that naturally arise in the exact definition of the renormalized effective action are not included in this formulation. These terms break dilatation invariance and are not accounted for by the scaleless action of Riegert-type. \\
The absence of scales in Riegert's analysis is quite obvious, since at $d=4$ no scales are generated by the Weyl rescaling of the metric. Therefore, the rescaling used by Riegert in the derivation of the anomaly induced action does not follow an ordinary DR procedure. \\ 
Our works \cite{Coriano:2022ftl,Coriano:2023sab} discuss these contributions by combining dimensional regularization (DR) with dimensional reduction (DRED) in order to establish a well-defined procedure for deriving the effective action at $d=4$ with the inclusion of logarithmic corrections associated both with the breaking of the dilatation symmetry for the dilatation anomaly $and$ the trace anomaly.\\
To regulate divergences in quantum field theory, dimensional regularization (DR) with finite renormalizations is commonly employed. However, there is ongoing debate over the optimal method to perform the $d\to 4$ limit, with a focus on addressing concerns arising from Weyl-invariant corrections that depend on the choice of extra dimensional metric and the regularization technique used. These corrections, absent in nonlocal actions obtained as functional solutions of the trace anomaly constraint, are significant in computing correlation functions evaluated around flat space.\\
It is important to note that $4d$ EGB theories are derived from the anomaly action by removing Weyl-invariant terms that result from quantum corrections. These corrections originate from the singular term $(1/\epsilon)V_{C^2}$, which is defined as the integral of the squared Weyl tensor $(C^2)$. This term is introduced at the 1-loop level in an anomaly action to regulate the contributions of such quantum corrections.
\\
The Weyl-invariant corrections are denoted as $\sm_f$ and $\tilde{\sm}_f$, and their specific form depends on the regularization method used. However, regardless of the choice of regularization, the corrections exhibit a logarithmic behavior. It is worth noting that both $V_{C^2}$ and $V_E$ are analytic in $d$, and as such, their formal expansions around $d=4$ are well-defined, at least for non-singular metrics.\\
Overall, the study of the relationship between anomaly actions and the $4d$ EGB theory provides important insights into the behavior of Weyl-invariant corrections and the most appropriate way to perform the $d\to 4$ limit.

\section{The conformal anomaly action}
To examine the backreaction of a conformal sector on the gravitational metric, we consider the partition function $\mathcal{Z}_B(g)$ given by the bare functional (in the Euclidean case) as

\begin{equation}
\label{partition}
\mathcal{Z}_B(g)=\mathcal{N}\int D\chi e^{-S_0(g,\chi)},
\end{equation}

where $\mathcal{N}$ is a normalization constant, and $\chi$ represents a conformal scalar. The 1-particle irreducible effective action is denoted by $\mathcal{S}_B(g)$, which is defined as the logarithm of the partition function:

\begin{equation}
\label{defg}
e^{-\mathcal{S}_B(g)}=\mathcal{Z}_B(g) \leftrightarrow \mathcal{S}_B(g)=-\log\mathcal{Z}_B(g).
\end{equation}

In this case, we assume that the quantum matter fields are in a conformal phase. The effective action $\mathcal{S}_B(g)$ includes all the multiple insertions of the stress energy tensor $T^{\mu\nu}_{scalar}$ given by
\bea
\label{defT}
T^{\mu\nu}_{scalar}
&\equiv&\frac{2}{\sqrt{g}}\frac{\delta S_0}{\delta g_{\mu\nu}}\nonumber \\
&=&\nabla^\mu \chi \, \nabla^\nu\chi - \frac{1}{2} \, g^{\mu\nu}\,g^{\alpha\beta}\,\nabla_\alpha \chi \, \nabla_\beta \chi
+ \chi \bigg[g^{\mu\nu} \Box - \nabla^\mu\,\nabla^\nu + \frac{1}{2}\,g^{\mu\nu}\,R - R^{\mu\nu} \bigg]\, \chi^2, 
\eea
which is renormalized using $V_E$ and $V_{C^2}$
in the form 
\begin{equation}
\mathcal{Z}_R(g)_=\, \mathcal{N}\int D\Phi e^{-S_0(g,\Phi) + b' \frac{1}{\epsilon}V_E(g,d) + b \frac{1}{\epsilon}V_{C^2}(g,d)},
\end{equation} 
where $\mathcal{N}$ is a normalization constant, and can be expressed in terms of stress energy tensor correlators $\langle T_1 T_2\ldots T_n\rangle$, with propagators and vertices that can be defined in any background, using \eqref{defT}.
The purpose of the two counterterms is to eliminate the divergences present in the bare effective action $\sm_B$, which is given by the expression
\begin{equation}
\sm_B(g,d)=-\log\left(\int D\Phi e^{-S(\Phi,g)}\right) +\log\mathcal{N}.
\end{equation}
This allows us to define a regularized effective action $\sm_R$, given by 
\begin{equation}
\label{rena}
\mathcal{S}_R(g,d)=\mathcal{S}_B(g,d) +  b' \frac{1}{\epsilon}V_E(g,d) + b \frac{1}{\epsilon}V_{C^2}(g,d),
\end{equation}
where $b'$ and $b$ are constants and $V_E$ and $V_{C^2}$ are integrals that depend on the choice of explicit metrics in the $d$-dimensional space. The expansion of these integrals needs to be performed with great care in order to properly eliminate the divergences in $\sm_B$.\\
We implement DR on the counterterms, and use the analiticity of the two functionals $V$ respect to $d$, expanding their expressions around $d=4$, to obtain
\begin{equation} 
\label{expand1}
V_{E/C^2}(g, d)=\left( V_{E/C^2}(g, 4) + \varepsilon 
V_{E/C^2}'(g,4) +O(\varepsilon^2) \right),
\end{equation}
in terms of only one background metric $(g)$. One can show by a Weyl rescaling that while $V_{C ^2}(g,4)=V_{C^2}(\bar g, 4)$ at order $\epsilon=d-4$ the the expansion in terms of $g$ and $\bar{g}$ 
differ by finite terms containing extra scales.  \\
Investigating the structure of \eqref{defg} requires consideration of the Weyl scalings in $d$ dimensions of $E$ and ${C^2}$, as well as their integrals $V_E$ and $V_{C^2}$. To analyze $\sm_B$ as $d\to 4$, we need to isolate the single pole divergences in $1/\epsilon$ that affect it. This can be achieved by expressing $\sm_B$ as

\begin{equation}
\label{pone}
\sm_B(g,d)= \lim_{d\to 4}\left(\sm_f(d) -\frac{b}{\epsilon}V_{C^2}(g,4) - \frac{b'}{\epsilon}V_E(g,4)\right),
\end{equation}
where $\sm_f$ is finite. The validity of \eqref{pone} is based on the fact that a conformal sector generates only singularities with a single pole in all the correlators, which can be canceled out by including $V_{C^2}$ along with the evanescent term $V_E$.\\
In $d$ dimensions, $\sm_B$ is both finite and Weyl-invariant. However, when we extract the singular contributions from $\sm_B$ and take the $d\to 4$ limit, we need to redefine the Weyl variation in a careful manner. Eq. \eqref{expand1} expands the residue at the $1/\epsilon$ pole for any background metric $g$, including the $O(\epsilon^0)$ terms $V_{E/C^2}(4)$ that require special treatment. In particular, the topological nature of $V_E(4)$ guarantees that it does not contribute to the infinite renormalization of the bare quantum action $\sm_B$ at $d=4$, as it is independent of any metric variation. In fact, $V'_E$ and its variants, defined through the Wess-Zumino part of $\sm_R$ in \eqref{rena}, provide finite renormalizations for all correlation functions of stress-energy tensors generated by the functional expansion of either $\sm_B$ or $\sm_f$ once the $d\to 4$ limit is taken. This behavior is analogous to the case of the chiral anomaly diagram, such as the AVV diagram, where the introduction of a Chern-Simons form preserves the vector Ward identities, and the vertex does not require an infinite counterterm to regulate its two form factors that are divergent by power-counting. Given that both the chiral anomaly and $V_E$ are topological, this similarity is not unexpected.\\
The renormalization procedure can be expressed by the equation
\beqa
\label{sum}
\sm_R(d)&=&\Big(\sm_f(d) -\frac{1}{\epsilon}V_{C^2}(g,4) - \frac{1}{\epsilon}V_E(g,4)\Big) \nn
&& +
\frac{1}{\varepsilon}\left( V_{E}(g,4) 
 + \varepsilon
V_{E}'(g,4) +O(\varepsilon^2) \right) + \frac{1}{\varepsilon}\left( V_{C^2}(g,4) + \varepsilon
V_{C^2}'(g,4) +O(\varepsilon^2) \right),
\eeqa
where $\sm_f$ is the finite action and $V_{C^2}(g,4)$ and $V_E(g,4)$ are introduced, respectively, to remove the $1/\epsilon$ singularities present in the bare effective action $\sm_B$ and to implement in the same action the Wess-Zumino consistency condition. After including the multiplicities $b$ and $b'$, that in free field theory realizations are proportional to the number of massless degrees of freedom of particles included in the loops, we obtain the expression
\begin{equation}
\label{sf1}
\sm_f(4)=\lim_{d\to 4}\Big(\sm_B(d) +\frac{b}{\epsilon}V_{C^2}(g,4) +\frac{b'}{\epsilon}V_E(g,4)\Big),
\end{equation}
which is Weyl invariant. It is important to note that the expansion of the counterterms in the second and third brackets of Eq. \eqref{sum} requires careful attention and explicit metrics in the $d$-dimensional integrals $V_{E/C^2}$.\\
It is also worth noting that in the expression for $\sm_R$, the $1/\epsilon$ contribution from $\sm_B$ cancels out with the expansions of the counterterms $V_{E/C^2}$. However, identifying these terms can be challenging due to the need to compute propagators and vertices in a curved background. In some cases, such computations can be performed using coordinate space methods in DR. For example, in the De Sitter case, point-splitting or other techniques can be used to efficiently compute 1-point functions such as $\langle T_{\mu\nu}\rangle$.
\section{Weyl rescalings and the local actions}
$V_E$ and $V_{C^2}$ are not Weyl invariant away from $d=4$. This will be important in identifying their dimensional expansions in $\epsilon$ around $d=4$. We will consider only the case in which $V_E$ is present, since we are not including any quantum correction in the EGB case, and assume that Greek indices run from 1 to $d$, unless specified otherwise. This discussion is crucial for highlighting the difference between the various ways in which the limit is performed, either using DR. For this we need to work out the response of $V_E$ to the Weyl rescaling of the metric, which is expressed in terms of a conformal factor (dilaton) $\phi$  and a fiducial metric $\bar{g}$. Notice that the rescaling 
\beq \label{weyiling transformazione finita} g_{\mu\nu} = e^{2 \phi(x)} \bar{g}_{\mu\nu}  \eeq 
allows to identify both $\bar{g}$ and $\phi$ as separate fields, whose equations of motion are related by a trace constraint \cite{Coriano:2023lmc}. This induces the relation 
\beq\label{form1}
\rg E=\sqrt{\bar g} e^{(d-4)\phi}\biggl \{ \bar E+(d-3)\bar\nabla_\mu \bar J^\mu(\bar{g},\phi) +(d-3)(d-4)\bar  K(\bar{g},\phi)  \biggl \},
\eeq
between the topological density evaluated in the metrics $g$ and $\bar{g}$ (the latter denoted as $\bar{E}$),
where we have defined
\begin{equation}\label{GBexJ}
\bar J^\mu(\bar{g},\phi)=8\bar R^{\mu\nu}\bar\nabla_\nu\phi-4\bar R\bar\nabla^\mu\phi+4(d-2)(\bar\nabla^\mu\phi\bar\Box\phi-\bar\nabla^\mu\bar\nabla^\nu\phi\bar\nabla_\nu\phi+\bar\nabla^\mu\phi\bar\nabla_\lambda\phi\bar\nabla^\lambda\phi),
\end{equation}
and
\begin{equation}\label{GBexK}
\bar K(\bar{g},\phi)=4\bar R^{\mu\nu}\bar\nabla_\mu\phi\bar\nabla_\nu\phi-2\bar R\bar\nabla_\lambda\phi\bar\nabla^\lambda\phi+4(d-2)\bar\Box\phi\bar\nabla_\lambda\phi\bar\nabla^\lambda\phi+(d-1)(d-2)(\bnabla_\lambda\phi\bnabla^\lambda\phi)^2
\end{equation}
which are evaluated in the fiducial metric. 
 The Weyl scaling allows to derive the relation
 
 \begin{equation} 
\label{ep2}
\frac{\delta}{\delta \phi}\int d^d y \sqrt{-g} E(y)=\epsilon \sqrt{g}E(x),
\end{equation} 
that can  be obtained either from \eqref{form1}, as specified above or, more simply, by a metric variation. In the second case one gets
\beq
\label{ep3}
 \dfun{}{g_{\mu \nu}} \int d^d x \rg E_4 = \rg \lt \frac{1}{2} g^{\mu \nu} E_4 - 2 R^{\mu \alpha \beta \gamma}R^\nu_{\alpha \beta \gamma} + 4 R^{\mu \alpha} R^\nu_{\ \alpha} + 4 R^{\mu \alpha \nu \beta} R_{\alpha \beta} - 2 R R^{\mu \nu} \rt  \eeq
and \eqref{ep2} follows if we contract with $2 g^{\mu\nu}$ both sides
\beq
\label{epx}
2 g_{\mu\nu}\frac{\delta}{\delta g_{\mu\nu}}\int d^d y \sqrt{-g} E(y)=\epsilon \sqrt{g}E(x).
\end{equation} 
This relation is true if we neglect boundary terms.  After renormalization, $\phi$ and $\bar{g}$ should be treated as independent fields, though constrained, as we have just mentioned, by a trace relation. Therefore, the identity 
\beq
2 g^{\mu\nu}\frac{\delta}{\delta g^{\mu\nu}}=\frac{\delta}{\delta\phi},
\eeq
which is used extensively in the computation of Weyl variations of a certain metric, includes only variations that keep $\bar{g}$ fixed. As discussed in \cite{Coriano:2022ftl}, after renormalization, one should deal with more general varations, where the dilaton and the fiducial metric should be treated as independent fields. This approach, if the subtractions are performed using a procedure which is pretty close to DR in flat space, does account for the breaking of the dilatation symmetry in a natural way by the conformal anomaly action \cite{Coriano:2023sab}. We are going to come back to this point in more detail below.\\
Coming  to the two possible definitions of the regulated effective action, the renormalized action $S_R$ can  be defined in two possible schemes 
 by the equations
 \beq
 \label{bone}
 \sm_R=\sm_f + \sm_A
 \eeq
 and
 \beq
 \label{btwo}
\sm_R=\tilde\sm_f + \sm_{WZ},
 \eeq
 where
 \beq
 \tilde\sm_f -\sm_{{f}}= \sm_A -\sm_{WZ}=\sm_{A/WZ}.
 \eeq
The finite action $\tilde\sm_f$ is identified by the relation
 \beq
\label{sf2}
\tilde\sm_f(4)=\lim_{d\to 4} \Big(\sm_B(d)  +\frac{1}{\epsilon}V_E(\bar g,d)\Big).
\eeq
In more detail, the renormalization, in the WZ case follows the pattern 
\beqa
\sm_R(d)&=&\Big( \sm_B(g,d)  +\frac{1}{\epsilon}V_E(\bar g,d)\Big) +
\frac{1}{\epsilon} \left( V_{E}(g,d) 
- V_{E}(\bar g,d)  \right), 
\eeqa
where 
the first term corresponds to $\tilde\sm_f$ while the second and the third contributions identify the WZ action 
\beq
\label{zz}
\mathcal{S}_{WZ}=\lim_{\epsilon\to 0}
\frac{1}{\epsilon} \left( V_{E}(g,d) 
- V_{E}(\bar g,d)  \right),
\eeq
corresponding to \eqref{bone}.
Notice that in this action, the limit is performed using both $g$ and $\bar{g}$ in $d$ dimensions. A different way to implement the subtractions is 
\beq
\mathcal{S}_{A}=\lim_{\epsilon\to 0}
\frac{1}{\epsilon} \left( V_{E}(g,d) 
- V_{E}( g,4)  \right),
\eeq
corresponding to \eqref{btwo}. In the case of the conformal anomaly action, where also $V_{C^2}$ is present, there are finite differences between the two regularizations. In particular, one can show that in this case the action will  contain a logarithm of the extra scales ($\log \mu L$) introduced by the regularization, limitedly to the $V_{C^2}$ counterterm. Details of this study can be found in \cite{Coriano:2023sab}.\\
In the case of the $4d$ EGB one considers only the contribution related to $V_E$ in \eqref{zz}, but the approach is, for the rest, identical to the one defined for the anomaly action. It has been used for a long time in the regularization of these theories. It has also been well-known that $V_E$ is not a counterterm, but, as mentioned, is needed in order to satisfy the WZ consistency condition in the Weyl variation of the anomaly functional.  \\
In the case of the $4d$ regulated EGB action, the GB contribution takes the form 
\beqa
\label{rdef}
\hat{V}'_E( g, \phi)&&\equiv \lim_{\epsilon\to 0}\left(\frac{1}{\epsilon}\left(V_E(g,d)-V_E(\bar g,d)\right)\right)\nn
&=&\int d^4x \rg \Big[\phi {}_4 E-(4 {} G^{\mu\nu}(\bar\nabla_\mu\phi\bar\nabla_\nu\phi)+2(\nabla_\lambda \phi \nabla^\lambda \phi )^2 +4\Box\phi \nabla_\lambda \phi \nabla^\lambda \phi ) \Big],
\eeqa
if we adopt the WZ definition, while in the other approach, corresponding to $\sm_{A}$, the subtraction is defined as 
\beq
\label{4d}
V'_E=\lim_{\epsilon\to 0}\left(\frac{1}{\epsilon}\left(V_E(g,d)-V_E(\bar g,4)\right)\right)
\eeq
with the two expressions differing by finite terms that depend on the regularization scales
\beq
V'_E=\hat{V}'_E(\bar g, \phi, d) +\log(L \mu)\int d^4x \sqrt{\bar g} \bar E.
\label{more}
\eeq
Notice that in the case of $V_E$ the log term does not play any role, since it is multiplied by a topological invariant factor, as clear from \eqref{more}. The derivation is worked out in the appendix to which we refer for more details.
The same is not true for the counterterm $V_{C^2}$. In this second case, if we perform a subtraction directly at $d=4$, similarly to \eqref{4d}, a log term is naturally present. Notice that if we perform a subtraction as in \eqref{4d}
logarithmic terms are naturally present. They are not part of the $4d$ GB theory but are part of the conformal anomaly action. This sets a distinction between anomalies of type "a" and "b" in a rather simple way, as discussed in \cite{Coriano:2023sab}, where the first ones are classified as topological and the second as non-topological. \\
 In the $4d$ EGB case, since the dilaton is part of the spectrum, a Weyl transformation needs to be performed also on the EH action beside the GB action, deriving the ordinary form of the dilaton gravity action
\bea 
\int d^d x \rg ( M_P^2 R-2 \Lambda) &=& \int d^dx \rgb \  e^{(d-2) \phi} \lt M_P^2 [_d\bar R - 2(d-1)\bar \square \phi - (d-1)(d-2) \bnabla_\lambda \phi \bnabla^\lambda \phi] - 2e^{2\phi} \Lambda \rt  \nn 
&& =\int d^dx \rgb  e^{(d-2) \phi} \left(M_P^2[\bar R + (d-1)(d-2) \bnabla_\lambda \phi \bnabla^\lambda \phi] - 2e^{2 \phi}\Lambda \rt, \eea
where $M_P$ is the Planck mass. The DRED of this action leads to the ordinary dilaton gravity $\sm_{EHd}$ in the Jordan frame
\beq
\label{d}
\sm_{EHd_1}(\bar g,\phi)=\int d^4x \rgb  e^{2 \phi} \left(M_P^2[\bar R + 6\bnabla_\lambda \phi \bnabla^\lambda \phi] - 2e^{2 \phi}\Lambda \rt. 
\eeq
It is obvious that a local EGB theory includes a dilaton both from the EH and the GB sectors. In the nonlocal version, in which the dilaton is removed by a finite renormalization of the topological term, the Weyl gauging of the EH action is not necessary. Both sectors are expressed only in terms of the metric $g$. We are now going to illustrate this point.   

\section{The nonlocal theory}
Let's discuss the nonlocal theory, where $V_E$ has distinct properties for $d \neq 4$. In  \cite{Riegert:1984kt} Riegert showed, using a Weyl rescaling, how to eliminate the dilaton in the conformal anomaly effective action, by inverting the linear relation for $\phi$ given by

\begin{equation}
\sqrt{g}\Big(E-\frac{2}{3}{\Box} R\Big)=
\sqrt{\bar g}\Big({\bar E}-\frac{2}{3}\,{\bar \Box} {\bar R}
+ 4{\bar \Delta_4}\phi \Big).
\label{119}
\end{equation}
Here, $\Delta_4$ is the fourth-order self-adjoint operator, which is conformal invariant when it acts on a scalar function of vanishing scaling dimensions
\beq
\Delta_4 = \nabla^2 + 2\,R^{\mu\nu}\nabla_\mu\nabla_\nu - \frac{2}{3}\,R{\Box}
+ \frac{1}{3}\,(\nabla^\mu R)\nabla_\mu, 
\label{120}
\eeq
and satisfies the relation
\beq
\sqrt{-g}\,\D_4\chi_0=\sqrt{-\bar g}\,\bar{\D}_4 \chi_0\label{point2},
\eeq
if $\chi_0$ is invariant (i.e. has scaling equal to zero) under a Weyl transformation. \\
Therefore, the regularization of the GB term can indeed generate regulated GB actions which can either take a local or a nonlocal form, depending on the way the conformal factor is treated in the regularization procedure \cite{Mazur:2001aa,Coriano:2022knl}.\\
Eq. \eqref{119} is crucial for the elimination of $\phi$ from the effective action. This is obtained by the 
inclusion of a boundary term $(\Box R)$. 
The method for identifying the anomaly action using Riegert's approach is different from the DR-based approaches discussed in the previous sections, which rely on appropriate metrics and manifolds of integrations. The 
approach, however, can be extended to $d$ dimensions, as discussed in \cite{Mazur:2001aa}, by including finite renormalizations to make it consistent with DR. In order to eliminate $\phi$, variants of $E$ that satisfy equation \eqref{ep2} can be introduced. An example is the modified and extended version of $E$, which includes a boundary contribution and an $O(\epsilon)$ modification. This term is useful for investigating the contribution of the $V_E$ counterterm to the effective action and plays a role in identifying a form of the effective action similar to Riegert's action. \\
The scaling relation \eqref{119} is unusual because its metric variation links boundary terms in the two metrics $g_{\mu\nu}$ and $\bar{g}_{\mu\nu}$. We use in $d=4$ the relations 

\beq
\label{intt}
\frac{1}{4}\delta (\sqrt{g} E)=\sqrt{g}\,\nabla_\sigma \delta X^\sigma,
\qquad 
\delta X^\sigma=\varepsilon^{\mu\nu\alpha \beta}\varepsilon^{\sigma\lambda\gamma\tau}
\delta \Gamma^\eta_{\nu\lambda}g_{\mu\eta}R_{\alpha \beta \gamma \tau},\qquad \varepsilon^{\mu\nu\alpha \beta}=\frac{\epsilon^{\mu\nu\alpha \beta}}{\sqrt{g}}
\eeq
under a metric variation $\delta$. These relations can be derived through integration by parts, while observing that the conformal factor behaves like a scalar under Weyl rescalings in two distinct frames $x$ and $x'$. Notice that a fiducial metric transforms as an ordinary tensor in these frames, and therefore, we have
\beq
g_{\mu\nu}(x)=\bar{g}_{\mu\nu}(x) e^{2 \phi(x)} \qquad g'_{\mu\nu}(x')=\bar{g}'_{\mu\nu}(x') e^{2 \phi'(x')}
\qquad \phi'(x')=\phi(x).
\eeq
We can define
\beq
\delta \Sigma^\sigma=\sqrt{g}g^{\sigma\beta}\partial_\beta \delta_\zeta \qquad
\label{int1}
\eeq
and vary both sides of \eqref{119} using the equation
\beq
\delta_\phi \left(\sqrt{g}\Box R\right) = \epsilon \delta\phi \Box R +(d-6)\sqrt{g}\nabla^\lambda R \nabla_\lambda \delta\phi -2 \sqrt{g}R\nabla^2\delta \phi -2 (d-1)\sqrt{g}\nabla^4\delta \phi
\eeq
which yields at $d=4$
\begin{align}
& \delta_\phi\left(\frac{1}{4}\sqrt{-g}\left(E-\frac{2}{3}\square R\right)\right)=\sqrt{- g}\D_4\delta\phi,
\label{pointd}
\end{align}
\beq
\partial_\sigma\left( \delta_\phi X^{\sigma} -\frac{1}{6}\delta_\phi\Sigma^\sigma\right)=\sqrt{- g}\D_4\delta\phi, 
\label{cc}
\eeq
if we use \eqref{intt} to relate it to a boundary contribution. 
Eq. \eqref{cc}, integrated over spacetime, gives consistently 
\beq
\int d^4 x \partial_\sigma\left( \delta_\phi X^{\sigma} -\frac{1}{6}\delta_\phi\Sigma^\sigma\right)=0,
\eeq
if we assume asymptotic flatness, and therefore
\beq
\int d^4 x \sqrt{- g}\D_4\delta\phi=0,
\eeq
that follows from the self-adjointness of $\Delta_4$
\begin{equation}
\int d^4x\sqrt{-g}\,\y(\D_4\x)=\int d^4x\sqrt{-g}\,(\D_4\y)\x\label{point3},
\end{equation}
where $\x$ and $\psi$ are scalar fields of zero scaling dimensions. Equation $\eqref{119}$ is only valid in the case where $d=4$ and is significantly simpler than the more general equation $\eqref{form1}$, which holds for any number of dimensions $d$. As we have already stressed, $\eqref{119}$ does not pertain directly to a DR procedure. \\
It is convenient to redefine \eqref{119} in the form 
\beq
 J(x)=\bar{J}(x) + 4 \sqrt{g}\Delta_4\phi(x),\qquad     \bar J(x)\equiv \sqrt{\bar g}\left( \bar E-\frac{2}{3}\bar \Box \bar R\right), \qquad  J(x)\equiv \sqrt{ g}\left(  E-\frac{2}{3} \Box  R\right) 
 \eeq
\begin{equation}
(\sqrt{-g}\,\D_4)_xD_4(x,y)=\d^4(x,y).\label{point4}
\end{equation}
and invert \eqref{119} using the properties of the operator $\Delta_4$ to identify $\phi(x)$
\begin{equation}
\label{onshell}
\phi(x)=\frac{1}{4}\int d^4y\,D_4(x,y)(J(y)- \bar{J}(y)).
\end{equation}
 The derivation of $\sm_{WZ}$ requires the solution of the equation 
\beq
\frac{\delta \mathcal{S}_{WZ}^{(GB)}}{\delta \phi}=J,
\eeq
that takes the form
\beq
\sm_{WZ}=\int d^4 x \sqrt{\bar g}\left(\bar J \phi + 2 \phi \Delta_4 \phi\right).
\eeq
Inserting the expression of $\phi$ \eqref{onshell} into this equation we obtain 
the WZ action given by
\beq
\sm_{WZ}(g)=\frac{1}{8}\int d^4 x d^4 y J(x) D_4(x,y) J(y),
\eeq
or
\begin{equation}
\mathcal{S}_{WZ}^{}(g) =\frac {1}{8}\!\int \!d^4x\sqrt{-g_x}\, \left(E - \frac{2}{3}\Box R\right)_{\!x} 
\int\! d^4x'\sqrt{-g_{x'}}\,D_4(x,x')\left(E - \frac{2}{3}\Box R \right)_{x'}.
\label{Snonl}
\end{equation}

In the context of DR and, in particular, in the analysis of the effective actions, it is clear that variants of the topological terms are possible.\\
The action above has been derived by using the properties of $E$ under Weyl rescaling at $d=4$, but it can be formulated in a consistent DR scheme, using \eqref{form1}. \\
Obviously, as we move away from $d=4$, modifications of such densities are possible.\\
 In general, we can modify such forms either by boundary terms, which play a role only if we include a spacetime boundary and/or by additional diffeomorphism invariant contributions of $O(\epsilon)$. \\
If we consider the modified expression of $V_E$ given by 

\beq
\tilde{V}_{E}=\int d^d x\sqrt{g} \left(E_4 + \epsilon\frac{R^2}{2 (d-1)^2}\right).
\eeq
with 
\beq
E_{ext}\equiv E_4 + \epsilon\frac{R^2}{2 (d-1)^2}
\eeq
and varying with respect to the metric one obtains
\beq
\delta_\phi\int d^d x \sqrt{g} E_{ext}=\epsilon \sqrt{g}(E_{ext}- \frac{2}{d-1} \Box R). 
\eeq
The expression above needs to be investigated with care in the $d\to 4$ limit since the metric is still $d$-dimensional. If we use DRED, then there will be a natural cutoff in the integral, $(\mu L)^\epsilon$ as in \eqref{more}, that can be consistently removed as $\epsilon\to 0$. 

\subsection{The nonlocal EGB expansion} 

The replacement of $V_E$ with $\tilde{V}_E$ is the key to reobtain the nonlocal action even in the presence of a generalized scaling relation such as the one presented in \eqref{form1}.
We redefine $S_{WZ}$
using $E_{ext}$ obtaining 
\beq
\tilde{V}_E=\int d^d x \sqrt{g}E_{ext}\, , 
 \eeq
and the action

\begin{equation}
\mathcal{S}^{(WZ)}_{GB} =\lim_{\epsilon\to 0}\frac{\alpha}{\epsilon}\left(\tilde{V}_E(\bar{g}_{\mu\nu}e^{2\phi},d)- \tilde{V}_E(\bar{g}_{\mu\nu},d\right).
\label{inter}
\eeq
To reobtain the nonlocal expression we can use the relation
\beq
\frac{\delta}{\delta\phi}\frac{1}{\epsilon}\tilde{V}_E(g_{\mu\nu},d)= \sqrt{g}\left(E-\frac{2}{3}\Box R +
\epsilon\frac{R^2}{2(d-1)^2}\right)
\eeq
in \eqref{inter} that gives
 \beqa
 \frac{\delta \mathcal{S}^{(WZ)}_{GB}}{\delta\phi}&=&\alpha\sqrt{g}\left(E-\frac{2}{3}\Box R \right)\nonumber \\
&=&\alpha\sqrt{\bar g}\left(\bar E-\frac{2}{3}\bar \Box\bar R + 4 \bar\Delta_4 \phi\right),
\label{solve}
\eeqa
and 
\begin{equation}
\mathcal{S}^{(WZ)}_{GB} = \alpha\int\,d^4x\,\sqrt{-\bar g}\,\left\{\left(\overline E - {2\over 3}
\bar{\Box} \overline R\right)\phi + 2\,\phi\bar\Delta_4\phi\right\}.\,
\label{WZ2}
\end{equation}

As before, we can solve for $\phi$, deriving the regulated GB action \eqref{Snonl}
that coincides with the result provided in \cite{Mazur:2001aa} by Mazur and Mottola if we remove in their result the contribution of the Weyl tensor $V_{C^2}$. \\
The nonlocal Einstein-Gauss-Bonnet (EGB) action can be expressed as a series expansion, particularly in the vicinity of a flat spacetime. This expansion is based on a dimensionless combination of the product of scalar curvature $R$ and the inverse of the D'Alembertian of flat space, referred to as $R\Box^{-1}$, which has been extensively investigated in perturbative computations and hierarchical analyses of the CWIs \cite{Coriano:2018bsy,Coriano:2017mux,Coriano:2021nvn}. By performing an expansion around flat space, it is possible to extract the classical interactions present in the action.
One rewrites the nonlocal anomaly action in an equivalent local form  
\beqa
\label{loc}
&&\hspace{-1.5cm} \mathcal{S}_{\rm anom}(g,\vf) \equiv -\sdfrac{1}{2} \int d^4x\,\sqrt{-g}\, \Big[ (\sq \vf)^2 - 2 \big(R^{\m\n} - \tfrac{1}{3} R g^{\m\n}\big)
(\nabla_\m\vf)(\nabla_\n \vf)\Big]\nn \\
&& \hspace{1.5cm} +\, \sdfrac{1}{2}\,\int d^4x\,\sqrt{-g}\  \Big[\big(E - \tfrac{2}{3}\sq R\big)  \Big]\,\phi,
\label{Sanom}
\eeqa
that can be varied with respect to $\phi$, giving
\be
\sqrt{-g}\,\D_4\, \vf = \sqrt{-g}\left[\sdfrac{E}{2}- \sdfrac{\!\sq R\!}{3} \right] \label{phieom}.
\ee
The metric and dilaton fields can be expressed perturbatively in the expansions 
\begin{align}
g_{\mu\nu} &= g_{\mu\nu}^{(0)} + g_{\mu\nu}^{(1)} + g_{\mu\nu}^{(2)} + \dots \equiv \eta_{\mu\nu} + h_{\mu\nu} + h_{\mu\nu}^{(2)} + \dots \\
\varphi &= \varphi^{(0)} + \varphi^{(1)} + \varphi^{(2)} + \dots
\end{align}

This expansion represents a set of terms obtained by considering $g_{\mu\nu} = \delta_{\mu\nu} + \kappa h_{\mu\nu}$, where $\kappa$ is the coupling expansion and $h$ has a mass dimension of one. Higher-order terms in the functional expansion of \eqref{loc}, of the order $h^2$, $h^3$, and so on, are collected. A similar expansion can be applied to $\varphi$, where $\varphi^{(1)}=\kappa \bar\varphi^{(1)}, \varphi^{(2)}=\kappa^2 \bar \varphi^{(2)}$, and so on. This results in the following relations
\beqa
\overline{\Box}^2 \vf^{(0)} &&= 0 \label{eom0}\\
(\sqrt{-g} \D_4)^{(1)} \vf^{(0)} + \overline{\Box}^2 \vf^{(1)} &&= \left[\sqrt{-g}
\left( \sdfrac{E}{2}- \sdfrac{\!\sq R\!}{3} \right)\right]^{(1)}
= - \sdfrac{\!1\!}{3}\, \overline{\Box} R^{(1)} \label{eom1}\\
(\sqrt{-g} \D_4)^{(2)} \vf^{(0)} + (\sqrt{-g} \D_4)^{(1)} \vf^{(1)} + \overline{\Box}^2 \vf^{(2)} &&=
\left[\sqrt{-g}\left(\sdfrac{E}{2}- \sdfrac{\!\sq R\!}{3}  \right)\right]^{(2)} \nn
&&= \sdfrac{1}{2}E^{(2)} - \sdfrac{1}{3}\, [\sqrt{-g}\sq R]^{(2)},  \label{eom2}
\eea

where $\overline{\Box}$ is the d'Alembert wave operator in flat Minkowski spacetime, and we have used the fact that $E$ is of second order in the fluctuations while the Ricci scalar $R$ starts at first order

\be
\vf^{(1)} = - \sdfrac{\!1 \!}{3\overline{\Box}}\, R^{(1)}
\label{vf1}
\ee
and the solution of (\ref{eom2}) is
\be
\vf^{(2)} = \sdfrac{1}{\overline{\Box}^2} \left\{ (\sqrt{-g} \D_4)^{(1)}\sdfrac{\!1 \!}{3\overline{\Box}} \, R^{(1)} 
+  \sdfrac{1}{2}E^{(2)} - \sdfrac{1}{3}\, [\sqrt{-g}\sq R]^{(2)} \right\}.
\label{vf2}
\ee
The quadratic term is given by
\be
\mathcal S_{\rm anom}^{(2)} = - \sdfrac{1}{2} \,\int d^4x \, \vf^{(1)} \overline{\Box}^2 \vf^{(1)} + \sdfrac{1}{2}\,\int d^4x \, \left( - \sdfrac{2}{3} \overline{\Box} R^{(1)}\right) \vf^{(1)}
= \sdfrac{1}{18} \,\int d^4x \, \left(R^{(1)}\right)^2.
\label{Sanom2}
\ee
Notice that all the $1/\Box$ terms cancel, giving a local contribution. 
The third order terms in the expansion are
\bea
\mathcal S_{\rm anom}^{(3)} &=&  - \sdfrac{1}{2} \int d^4x \, \left\{2\,\vf^{(1)} \overline{\Box}^2 \vf^{(2)} +\vf^{(1)} \big(\sqrt{-g} \D_4\big)^{(1)} \,\vf^{\!(1)} \right\}\nn
&&\hspace{-1cm} + \sdfrac{1}{2} \int d^4x \left\{\left( - \sdfrac{2}{3} \overline{\Box} R^{(1)}\right) \vf^{(2)} + \left(E^{(2)} - \sdfrac{2}{3}\, \sqrt{-g}\sq R\right)^{\!(2)} \vf^{(1)} \right\}.
\label{Sanom3a}
 \eea
The remaining terms in (\ref{Sanom3a}) give
\bea
&&\mathcal S_{\rm anom}^{(3)} =- \sdfrac{1}{18} \int d^4x \, \left\{R^{(1)}\sdfrac{1}{\overline{\Box}} \big(\sqrt{-g} \D_4\big)^{\!(1)} \,\sdfrac{1}{\overline{\Box}} R^{(1)} \right\}
- \sdfrac{b'}{6} \int d^4x \left(E- \sdfrac{2}{3}\, \sqrt{-g}\sq R\right)^{\!(2)}\,\sdfrac{1}{\overline{\Box}} R^{(1)}.\nn
&&\hspace{2cm}. 
\eea
If we rearrange $\Delta_4$ in the form
\beq
\Delta_4= \Box^2 +2 \nabla_\mu(R^{\mu\nu}\nabla^\nu) -\frac{2}{3}\nabla_\mu(R \nabla^\mu),
\eeq
an expansion of this operator to first order in $\delta g_{\mu\nu}$ gives

\be
\big(\sqrt{-g} \D_4\big)^{\!(1)} =  \big(\sqrt{-g} \sq^2\big)^{\!(1)} + 2\, \pa_{\m} \left(R^{\m\n} - \sdfrac{1}{3} \eta^{\m\n} R\right)^{\!(1)}\pa_{\n}.
\ee
An integration by parts allows to extract the vertex describing the classical interaction of three gravitational waves 
\bea
&&\hspace{-1.1cm}\mathcal S_{\rm anom}^{(3)}\! =\!- \sdfrac{1}{18}\! \int\! d^4x \left\{\!R^{(1)}\!\sdfrac{1}{\overline{\Box}} \big(\sqrt{-g} \sq^2\big)^{\!(1)}\sdfrac{1}{\overline{\Box}} R^{(1)}\! \right\}
+ \sdfrac{1}{9}\! \int\! d^4x \left\{\!\pa_{\m} R^{(1)}\!\sdfrac{1}{\overline{\Box}} \! \left(\!R^{(1)\m\n}\! - \!\sdfrac{1}{3} \eta^{\m\n} R^{(1)}\!\right)\!
\sdfrac{1}{\overline{\Box}}\pa_{\n} R^{(1)}\!\right\}\hspace{-5mm}\nn 
&&\hspace{-8mm} - \sdfrac{1}{6}\! \int\! d^4x  E^{\!(2)} \sdfrac{1}{\overline{\Box}}R^{(1)}
+ \sdfrac{1}{9} \!\int\! d^4x\,  R^{(1)}  \sdfrac{1}{\overline{\Box}} \left(\sqrt{-g}\sq\right)^{\!(1)}R^{(1)}
+ \sdfrac{1}{9} \! \int\! d^4x\, R^{\!(2)}R^{(1)},
\label{Sanom3b}
\eea
characterised just by single propagator poles if transformed to momentum space. The evaluation of the expansion requires a set of operatorial identities in the flat limit 

\beq
\left(\sqrt{g}\Box^2\right)^{(1)}=-(\sqrt{g})^{(1)}\overline{\Box}^2 + (\sqrt{g}\Box)^{(1)}\overline{\Box} + \overline{\Box}(\sqrt{g}\Box)^{(1)},
\eeq 
to derive the nonlocal expression of the GB term
\bea
&&\hspace{-5mm} \mathcal S_{\rm anom}^{(3)} =
 \sdfrac{1}{9} \int\! d^4x \int\!d^4x'\!\int\!d^4x''\!\left\{\big(\pa_{\m} R^{(1)})_x\left(\sdfrac{1}{\overline{\Box}}\right)_{\!xx'}  
 \!\left(R^{(1)\m\n}\! - \!\sdfrac{1}{3} \eta^{\m\n} R^{(1)}\right)_{x'}\!
\left(\sdfrac{1}{\overline{\Box}}\right)_{\!x'x''}\!\big(\pa_{\n} R^{(1)})_{x''}\right\}\nn
&&\hspace{-6mm}- \sdfrac{1}{6}\! \int\! d^4x\! \int\!d^4x'\! \left(\, E^{\!(2)}\right)_{\!x}\! \left(\sdfrac{1}{\overline{\Box}}\right)_{\!xx'} \!R^{(1)}_{x'}
 + \sdfrac{1}{18} \! \int\! d^4x\, R^{(1)}\left(2\, R^{\!(2)} + (\sqrt{-g})^{(1)} R^{(1)}\right),
\label{S3anom3}
\eea
where the last term is purely local. This action can be used to investigate the graviton interactions up to trilinear perturbations around flat space. The expansion of these actions up to quartic order \eqref{Sanom} requires a separate study since it has been shown that the action fails to reproduce the perturbative expansion at quartic order. \\
 The $R\Box^{-1}$ operator appearing in the expansion provides a link with the 
 nonlocal cosmological models for dark energy \cite{Capozziello:2021krv}. Previous analysis around flat space have shown \cite{Giannotti:2008cv,Armillis:2009pq,Coriano:2018zdo} for 3-point functions in the $TJJ$ case and in \cite{Coriano:2017mux} for the 3-graviton vertex TTT that the action is consistently defined. Investigations of this action in the case of 4-graviton vertices have been presented in \cite{Coriano:2023lmc}  to which we refer for further details.
\section{Conclusions} 
The ordinary Einstein-Hilbert (EH) action with a gravitational constant is commonly used to describe the evolution of the universe. However, this approach may not provide a complete picture, as modifications to General Relativity (GR) such as $R^2$ or $f(R)$ theories, and topological corrections to GR, may also play a crucial role. Advancements in gravitational wave detectors may enable us to test these modifications and gain a deeper understanding of the nature of our universe.\\
One class of modified gravity theories that has attracted attention is the $4d$ Einstein-Gauss-Bonnet (EGB) theories, which can be obtained through a singular limit of the Euler-Poincar\`e density. This method is akin to the construction of conformal anomaly actions, which arise from conformal symmetry, quantum corrections, and their breaking due to the conformal anomaly. However, in the case of $4d$ EGB theories, the approach is purely geometric and mathematical, enabling the inclusion of topological terms in the form of dilaton gravity models. The  outcome depends on how the dilaton field is extracted from the fiducial metric and on the treatment of its dependence on the extra dimensions.\\
Our work shows that nonlocal versions of these theories are feasible, drawing upon previous analyses of conformal anomaly actions. $R\Box^{-1}$ theories, considered as possible modifications of GR that may account for dark energy in cosmology, find a natural context within the formulation of the nonlocal conformal anomaly action, where their $1/\Box$ behaviour is associated with the exchange of conformal anomaly poles. The general feature of anomalies is the presence of massless exchanges 
 \\
The fundamental idea in favour of such formulations is that conformal symmetry in the early universe, recognized as 
an important pillar of physics at the Planck scale, can be broken by quantum corrections and can account for the dark energy. $4d$ EGB carry similar, though far more limited tracts of such anomaly actions. They are classical variants of such mother theories. 


\centerline{\bf Acknowledgements}
The work of M.C. is supported by a PON fellowship and partially supported by the Italian Institute of Technology. 
The work of C. C. and S.L. is funded by the European Union, Next Generation EU, PNRR project "National
Centre for HPC, Big Data and Quantum Computing", project code CN00000013 and by INFN
iniziativa specifica QGSKY.
M. M. M. is supported by the European Research Council (ERC) under the European Union as Horizon 2020 research and innovation
program (grant agreement No818066) and by Deutsche Forschungsgemeinschaft (DFG, German Research Foundation) under Germany's Excellence Strategy EXC-2181/1 - 390900948 (the Heidelberg
STRUCTURES Cluster of Excellence).

\appendix
\section{Cutoffs from the extra dimensions} 
We illustrate in this appendix how cutoffs emerge in the analysis of these actions in DRED. 
We use the Weyl gauging procedure to expand the $V_E$ counterterm. 
We can use \eqref{form1} to write the $d$- dimensional $V_E$  as
\beq V_E(d)= \mu^\epsilon\int d^d x \rg \ E = \mu^\epsilon\int d^d x \rg \ e^{(d-4)\phi}  [\bar E + (d-3) \bnabla_\mu \bar J^\mu + (d-3)(d-4) \bar K] , \eeq
We expand in series near $d=4$ the exponential $e^{(d-4)\phi}$ inside the integral

\bea
V_E(d)&=& \mu^\epsilon\int d^d x \rgb \ e^{(d-4)\phi}[\bar E+(d-3)\bar\nabla_\mu \bar J^\mu+(d-3)(d-4)\bar K]=\nn
&&=\sum_{n=0}^\infty \frac{(d-4)^n}{n!}\mu^\epsilon\int d^dx \sqrt{\bar g}[\phi^n \bar E+(d-3)(\phi^n\bar\nabla_\mu \bar  J^\mu+n\phi^{n-1} \bar K)].
\eea
Using the above relation, the counterterm reads
\begin{align}
&\frac{1}{d-4}V_E(d)=\frac{1}{(d-4)}\sum_{n=0}^\infty \frac{(d-4)^n}{n!}\mu^\epsilon\int d^dx \rgb \ [\phi^n \bar E+(d-3)(\phi^n\bar\nabla_\mu \bar J^\mu+n\phi^{n-1} \bar K)].
\end{align}
An expansion up to $O(1)$ in $\epsilon$ gives
\begin{align}
&\frac{1}{d-4}V_E(d)=\frac{\mu^\epsilon}{(d-4)}\int d^dx \rgb\ [\bar E+(\bar\nabla_\mu \bar J^\mu)]+\mu^\epsilon\int d^dx \rgb \ [\phi\bar E+(\phi\bar\nabla_\mu \bar J^\mu+ \bar K)] + O(d-4).
\end{align}
We can neglect the $\bar\nabla_\mu \bar J^\mu$ in the first term, since it is a boundary term and integrate  by parts $\phi\bar\nabla_\mu \bar J^\mu$, obtaining 
\begin{align}
&\frac{\mu^\epsilon}{d-4}V_E(d)=\frac{\mu^\epsilon}{(d-4)}\int d^dx \rgb\ \bar E+ \mu^\epsilon\int d^dx \rgb \ [\phi\bar E+(\phi\bar\nabla_\mu \bar J^\mu+ \bar K)] + O(d-4).
\end{align}
Using \eqref{GBexJ} and \eqref{GBexK} we can write
$$
\bar \nabla_\mu\phi \bar J^\mu-\bar K=4\bar R^{\mu\nu}(\bnabla_\mu\phi\bnabla_\nu\phi)-2\bar R\bar \square \phi+2(\bnabla_\lambda \phi \bnabla^\lambda \phi )^2+4\bar\Box\phi\bnabla_\lambda \phi \bnabla^\lambda \phi ,
$$
and after an integration by parts we get the final form of the counterterm up to $O(d-4)$ terms
\bea 
\frac{1}{d-4}V_E(d) &=& \frac{\mu^\epsilon}{(d-4)}\int d^dx \rgb \ \bar E+ \mu^\epsilon\int d^dx \rgb\ \Big[\phi\bar E-(4\bar G^{\mu\nu}(\bar\nabla_\mu\phi\bar\nabla_\nu\phi) \nn && +2(\bnabla_\lambda \phi \bnabla^\lambda \phi )^2 +4\bar\Box\phi \bnabla_\lambda \phi \bnabla^\lambda \phi ) \Big].
\eea
If we use DRED and neglectg the dependence of $\phi(x)$ on the extra dimensions, the integration in the extra dimensions is performed trivially assuming a space cutoff $L$ and obtain  
\bea \label{WG in generic case}
\frac{1}{d-4}V_E(d) &=&  \frac{(\mu L)^{(d-4)}}{(d-4)} \int d^4x \rg \ \bar E+ (\mu L)^{(d-4)} \int d^4 x \rgb\ \Big[\phi\bar E-(4\bar G^{\mu\nu}(\bar\nabla_\mu\phi\bar\nabla_\nu\phi) \nn && +2(\bnabla_\lambda \phi \bnabla^\lambda \phi )^2 +4\bar\Box\phi \bnabla_\lambda \phi \bnabla^\lambda \phi ) \Big],
\eea
where $L^{(d-4)}$ is the volume of the extra dimensional space. 
In the case of the GB term we can send $\epsilon\to 0$ and observe that the log term, in this limit, is multiplied by a the value of $V_E$ at $d=4$, which is a topological invariant, as shown in \eqref{more}. Therefore no log terms are generated.


\begin{thebibliography}{10}

\bibitem{Stelle:1977ry}
K.~S. Stelle, {\it {Classical Gravity with Higher Derivatives}},  {\em Gen.
  Rel. Grav.} {\bf 9} (1978) 353--371.

\bibitem{Zwiebach:1985uq}
B.~Zwiebach, {\it {Curvature Squared Terms and String Theories}},  {\em Phys.
  Lett. B} {\bf 156} (1985) 315--317.

\bibitem{Starobinsky:1980te}
A.~A. Starobinsky, {\it {A new type of isotropic cosmological models without
  singularity}},  {\em Phys. Lett.} {\bf B91} (1980) 99--102.

\bibitem{Antoniadis:2020dfq}
I.~Antoniadis, A.~Lykkas, and K.~Tamvakis, {\it {Constant-roll in the
  Palatini-$R^2$ models}},  {\em JCAP} {\bf 04} (2020), no.~04 033,
  [\href{http://xxx.lanl.gov/abs/2002.1268}{{\tt arXiv:2002.1268}}].

\bibitem{Capozziello:2021krv}
S.~Capozziello and F.~Bajardi, {\it {Nonlocal gravity cosmology: An overview}},
   {\em Int. J. Mod. Phys. D} {\bf 31} (2022), no.~06 2230009,
  [\href{http://xxx.lanl.gov/abs/2201.0451}{{\tt arXiv:2201.0451}}].

\bibitem{Kanti:2015pda}
P.~Kanti, R.~Gannouji, and N.~Dadhich, {\it {Gauss-Bonnet Inflation}},  {\em
  Phys. Rev. D} {\bf 92} (2015), no.~4 041302,
  [\href{http://xxx.lanl.gov/abs/1503.0157}{{\tt arXiv:1503.0157}}].

\bibitem{Lovelock:1971yv}
D.~Lovelock, {\it {The Einstein tensor and its generalizations}},  {\em J.
  Math. Phys.} {\bf 12} (1971) 498--501.

\bibitem{Charmousis:2014mia}
C.~Charmousis, {\it {From Lovelock to Horndeski`s Generalized Scalar Tensor
  Theory}},  {\em Lect. Notes Phys.} {\bf 892} (2015) 25--56,
  [\href{http://xxx.lanl.gov/abs/1405.1612}{{\tt arXiv:1405.1612}}].

\bibitem{Lanczos:1938sf}
C.~Lanczos, {\it {A Remarkable property of the Riemann-Christoffel tensor in
  four dimensions}},  {\em Annals Math.} {\bf 39} (1938) 842--850.

\bibitem{Lanczos:1932zz}
C.~Lanczos, {\it {Electricity as a natural property of Riemannian geometry}},
  {\em Rev. Mod. Phys.} {\bf 39} (1932) 716--736.

\bibitem{Woodard:2015zca}
R.~P. Woodard, {\it {Ostrogradsky's theorem on Hamiltonian instability}},  {\em
  Scholarpedia} {\bf 10} (2015), no.~8 32243,
  [\href{http://xxx.lanl.gov/abs/1506.0221}{{\tt arXiv:1506.0221}}].

\bibitem{Glavan:2019inb}
D.~Glavan and C.~Lin, {\it {Einstein-Gauss-Bonnet Gravity in Four-Dimensional
  Spacetime}},  {\em Phys. Rev. Lett.} {\bf 124} (2020), no.~8 081301,
  [\href{http://xxx.lanl.gov/abs/1905.0360}{{\tt arXiv:1905.0360}}].

\bibitem{Lu:2020iav}
H.~Lu and Y.~Pang, {\it {Horndeski gravity as $D \rightarrow 4$ limit of
  Gauss-Bonnet}},  {\em Phys. Lett. B} {\bf 809} (2020) 135717,
  [\href{http://xxx.lanl.gov/abs/2003.1155}{{\tt arXiv:2003.1155}}].

\bibitem{Matsumoto:2022fln}
M.~Matsumoto and Y.~Nakayama, {\it {Dilaton invading from infinitesimal extra
  dimension}},  \href{http://xxx.lanl.gov/abs/2202.1353}{{\tt
  arXiv:2202.1353}}.

\bibitem{Aoki:2020lig}
K.~Aoki, M.~A. Gorji, and S.~Mukohyama, {\it {A consistent theory of $D \to 4$
  Einstein-Gauss-Bonnet gravity}},  {\em Phys. Lett. B} {\bf 810} (2020)
  135843, [\href{http://xxx.lanl.gov/abs/2005.0385}{{\tt arXiv:2005.0385}}].

\bibitem{Hennigar:2020lsl}
R.~A. Hennigar, D.~Kubiz\v{n}\'ak, R.~B. Mann, and C.~Pollack, {\it {On taking
  the D to 4 limit of Gauss-Bonnet gravity: theory and solutions}},  {\em JHEP}
  {\bf 07} (2020) 027, [\href{http://xxx.lanl.gov/abs/2004.0947}{{\tt
  arXiv:2004.0947}}].

\bibitem{Gurses:2020ofy}
G.~Metin, T.~C. Sisman, and T.~Bayram, {\it {Is there a novel
  Einstein\textendash{}Gauss\textendash{}Bonnet theory in four dimensions?}},
  {\em Eur. Phys. J. C} {\bf 80} (2020), no.~7 647,
  [\href{http://xxx.lanl.gov/abs/2004.0339}{{\tt arXiv:2004.0339}}].

\bibitem{Coriano:2022ftl}
C.~Corian\`o, M.~M. Maglio, and D.~Theofilopoulos, {\it {Topological
  corrections and conformal backreaction in the Einstein
  Gauss\textendash{}Bonnet/Weyl theories of gravity at $D=4$}},  {\em Eur.
  Phys. J. C} {\bf 82} (2022), no.~12 1121,
  [\href{http://xxx.lanl.gov/abs/2203.0421}{{\tt arXiv:2203.0421}}].

\bibitem{Mazur:2001aa}
P.~O. Mazur and E.~Mottola, {\it {Weyl cohomology and the effective action for
  conformal anomalies}},  {\em Phys.Rev.} {\bf D64} (2001) 104022,
  [\href{http://xxx.lanl.gov/abs/hep-th/0106151}{{\tt hep-th/0106151}}].

\bibitem{Codello:2012sn}
A.~Codello, G.~D'Odorico, C.~Pagani, and R.~Percacci, {\it {The Renormalization
  Group and Weyl-invariance}},  {\em Class.Quant.Grav.} {\bf 30} (2013) 115015,
  [\href{http://xxx.lanl.gov/abs/1210.3284}{{\tt arXiv:1210.3284}}].

\bibitem{Coriano:2013xua}
C.~Corian\`o, L.~Delle~Rose, C.~Marzo, and M.~Serino, {\it {Conformal Trace
  Relations from the Dilaton Wess-Zumino Action}},  {\em Phys. Lett. B} {\bf
  726} (2013), no.~4-5 896--905, [\href{http://xxx.lanl.gov/abs/1306.4248}{{\tt
  arXiv:1306.4248}}].

\bibitem{Coriano:2013nja}
C.~Corian\`o, L.~Delle~Rose, C.~Marzo, and M.~Serino, {\it {The dilaton
  Wess-Zumino action in six dimensions from Weyl gauging: local anomalies and
  trace relations}},  {\em Class. Quant. Grav.} {\bf 31} (2014) 105009,
  [\href{http://xxx.lanl.gov/abs/1311.1804}{{\tt arXiv:1311.1804}}].

\bibitem{Coriano:2023lmc}
C.~Corian\`o, M.~Creti, S.~Lionetti, and M.~M. Maglio, {\it {Three-Wave and
  Four-Wave Interactions in the $4d$ Einstein Gauss-Bonnet (EGB) and Lovelock
  Theories}},  \href{http://xxx.lanl.gov/abs/2302.0210}{{\tt arXiv:2302.0210}}.

\bibitem{Coriano:2023sab}
C.~Corian\`o and M.~M. Creti, Mario an d~Maglio, {\it {Broken Scale Invariance
  and the Regularization of a Conformal Sector in Gravity with Wess-Zumino
  actions}},  \href{http://xxx.lanl.gov/abs/2301.0746}{{\tt arXiv:2301.0746}}.

\bibitem{Coriano:2022jkn}
C.~Corian\`o, M.~M. Maglio, and R.~Tommasi, {\it {Four-point functions of
  gravitons and conserved currents of CFT in momentum space: testing the
  nonlocal action with the TTJJ}},
  \href{http://xxx.lanl.gov/abs/2212.1277}{{\tt arXiv:2212.1277}}.

\bibitem{Belgacem:2017cqo}
E.~Belgacem, Y.~Dirian, S.~Foffa, and M.~Maggiore, {\it {Nonlocal gravity.
  Conceptual aspects and cosmological predictions}},  {\em JCAP} {\bf 03}
  (2018) 002, [\href{http://xxx.lanl.gov/abs/1712.0706}{{\tt
  arXiv:1712.0706}}].

\bibitem{Capozziello:2021bki}
S.~Capozziello and M.~Capriolo, {\it {Gravitational waves in non-local
  gravity}},  {\em Class. Quant. Grav.} {\bf 38} (2021), no.~17 175008,
  [\href{http://xxx.lanl.gov/abs/2107.0697}{{\tt arXiv:2107.0697}}].

\bibitem{Edgar:2001vv}
S.~B. Edgar and A.~Hoglund, {\it {Dimensionally dependent tensor identities by
  double antisymmetrization}},  {\em J. Math. Phys.} {\bf 43} (2002) 659--677,
  [\href{http://xxx.lanl.gov/abs/gr-qc/0105066}{{\tt gr-qc/0105066}}].

\bibitem{lovelock_1970}
D.~Lovelock, {\it Dimensionally dependent identities},  {\em Mathematical
  Proceedings of the Cambridge Philosophical Society} {\bf 68} (1970), no.~2
  345?350.

\bibitem{Fernandes:2020nbq}
P.~G.~S. Fernandes, P.~Carrilho, T.~Clifton, and D.~J. Mulryne, {\it
  {Derivation of Regularized Field Equations for the Einstein-Gauss-Bonnet
  Theory in Four Dimensions}},  {\em Phys. Rev. D} {\bf 102} (2020), no.~2
  024025, [\href{http://xxx.lanl.gov/abs/2004.0836}{{\tt arXiv:2004.0836}}].

\bibitem{Fernandes:2022zrq}
P.~G.~S. Fernandes, P.~Carrilho, T.~Clifton, and D.~J. Mulryne, {\it {The 4D
  Einstein\textendash{}Gauss\textendash{}Bonnet theory of gravity: a review}},
  {\em Class. Quant. Grav.} {\bf 39} (2022), no.~6 063001,
  [\href{http://xxx.lanl.gov/abs/2202.1390}{{\tt arXiv:2202.1390}}].

\bibitem{Riegert:1984kt}
R.~J. Riegert, {\it {A Nonlocal Action for the Trace Anomaly}},  {\em Phys.
  Lett.} {\bf 134B} (1984) 56--60.

\bibitem{Coriano:2022knl}
C.~Corian\`o and M.~M. Maglio, {\it {Einstein Gauss-Bonnet theories as
  ordinary, Wess-Zumino conformal anomaly actions}},  {\em Phys. Lett. B} {\bf
  828} (2022) 137020, [\href{http://xxx.lanl.gov/abs/2201.0751}{{\tt
  arXiv:2201.0751}}].

\bibitem{Coriano:2018bsy}
C.~Corian\`o and M.~M. Maglio, {\it {The general 3-graviton vertex ($TTT$) of
  conformal field theories in momentum space in $d =4$}},  {\em Nucl. Phys.}
  {\bf B937} (2018) 56--134, [\href{http://xxx.lanl.gov/abs/1808.1022}{{\tt
  arXiv:1808.1022}}].

\bibitem{Coriano:2017mux}
C.~Corian\`o, M.~M. Maglio, and E.~Mottola, {\it {TTT in CFT: Trace Identities
  and the Conformal Anomaly Effective Action}},  {\em Nucl. Phys.} {\bf B942}
  (2019) 303--328, [\href{http://xxx.lanl.gov/abs/1703.0886}{{\tt
  arXiv:1703.0886}}].

\bibitem{Coriano:2021nvn}
C.~Corian\`o, M.~M. Maglio, and D.~Theofilopoulos, {\it {The Conformal Anomaly
  Action to Fourth Order (4T) in $d=4$ in Momentum Space}},
  \href{http://xxx.lanl.gov/abs/2103.1395}{{\tt arXiv:2103.1395}}.

\bibitem{Giannotti:2008cv}
M.~Giannotti and E.~Mottola, {\it {The Trace Anomaly and Massless Scalar
  Degrees of Freedom in Gravity}},  {\em Phys. Rev.} {\bf D79} (2009) 045014,
  [\href{http://xxx.lanl.gov/abs/0812.0351}{{\tt arXiv:0812.0351}}].

\bibitem{Armillis:2009pq}
R.~Armillis, C.~Corian\`{o}, and L.~Delle~Rose, {\it {Conformal Anomalies and
  the Gravitational Effective Action: The $TJJ$ Correlator for a Dirac
  Fermion}},  {\em Phys. Rev.} {\bf D81} (2010) 085001,
  [\href{http://xxx.lanl.gov/abs/0910.3381}{{\tt arXiv:0910.3381}}].

\bibitem{Coriano:2018zdo}
C.~Corian\`o and M.~M. Maglio, {\it {Renormalization, Conformal Ward Identities
  and the Origin of a Conformal Anomaly Pole}},  {\em Phys. Lett.} {\bf B781}
  (2018) 283--289, [\href{http://xxx.lanl.gov/abs/1802.0150}{{\tt
  arXiv:1802.0150}}].

\end{thebibliography}


\providecommand{\href}[2]{#2}\begingroup\raggedright\endgroup

\end{document}